\documentclass[12pt,letterpaper]{article}
\pdfoutput=1

\usepackage{graphicx,array}
\usepackage{color}
\usepackage{latexsym}
\usepackage{amsthm}
\usepackage{amsmath}
\usepackage{amssymb}
\usepackage[numbers,sort&compress]{natbib}
\usepackage{bm}
\usepackage{mathrsfs}
\usepackage{tensor}
\usepackage{mathtools}
\usepackage{relsize}
 
\usepackage{hyperref} %Automatically links \label and \ref commands; Always load last
\hypersetup{
    colorlinks=true,       % false: boxed links; true: colored links
    linkcolor=blue,        % color of internal links
    citecolor=blue,        % color of links to bibliography
    filecolor=magenta,     % color of file links
    urlcolor=blue          % color of external links
}
\usepackage[all]{hypcap} %Link navagates to top of figure instead of caption (below fig)

\numberwithin{equation}{section}

\newcommand{\rref}[1]{Ref.~\cite{#1}}
\newcommand{\fref}[1]{Fig.~\ref{#1}}
\newcommand{\Ccal}{\mathcal{C}}
\newcommand{\ud}{\mathrm{d}}
\newcommand{\abs}[1]{\left| #1 \right|}
\newcommand{\xvec}{{\bm x}}
\newcommand{\per}{\ . \ }
\newcommand{\com}{\ , \ }
\newcommand{\Mpl}{M_{\rm pl}}
\newcommand{\GeV}{\ \mathrm{GeV}}
\newcommand{\kvec}{{\bm k}}
\newcommand{\pref}[1]{(\ref{#1})}
\newcommand{\km}{\ \mathrm{km}}
\newcommand{\meV}{\ \mathrm{meV}}
\newcommand{\eref}[1]{Eq.~(\ref{#1})}
\newcommand{\sref}[1]{Sec.~\ref{#1}}
\newcommand{\nn}{\nonumber \\}
\newcommand{\const}{\, \mathrm{const.}}
\renewcommand{\exp}{\mathrm{exp}}
\newcommand{\IR}{\text{\sc \fontfamily{ptm}\selectfont ir}}
\newcommand{\slow}{\mathrm{slow}}
\newcommand{\fast}{\mathrm{fast}}
\newcommand{\Cxi}{\frac{1}{6} (1-6\xi)}
\newcommand{\Hinf}{H_\mathrm{inf}}
\newcommand{\phiC}{\phi_{\text{\sc \fontfamily{ptm}\selectfont c}}}
\newcommand{\dphiC}{\dot{\phi}_{\text{\sc \fontfamily{ptm}\selectfont c}}}
\newcommand{\tphiC}{\tilde{\phi}_{\text{\sc \fontfamily{ptm}\selectfont c}}}
\newcommand{\tend}{{t_\mathrm{end}}}
\newcommand{\Cc}{\Ccal_\mathrm{c}}
\newcommand{\Cm}{\Ccal_\mathrm{m}}

\title{\bf Gravitational production of super-Hubble-mass particles: an analytic approach}

\author{\large Daniel J. H. Chung$^{a}$, \ Edward W. Kolb$^{b}$, \ and \ Andrew J. Long$^{c}$}
\date{\small \it 
$^a$Department of Physics, University of Wisconsin-Madison, Madison,
WI 53706, USA \\ 
$^b$Kavli Institute for Cosmological Physics and Enrico Fermi Institute, University of Chicago, Chicago, IL 60637, USA \\
$^c$Leinweber Center for Theoretical Physics, University of Michigan, Ann Arbor, MI 48109, USA 
}

\begin{document}

\maketitle

\begin{abstract}

Through a mechanism similar to perturbative particle scattering, particles of mass $m_\chi$ larger than the Hubble expansion rate $H_\mathrm{inf}$ during inflation can be gravitationally produced at the end of inflation without the exponential suppression powers of $\exp(-m_\chi/H_\mathrm{inf})$.  Here we develop an analytic formalism for computing particle production for such massive particles. We apply our formalism to specific models that have been previously been studied only numerically, and we find that our analytical approximations reproduce those numerical estimates well.  
\end{abstract}

\begingroup
\hypersetup{linkcolor=black}
\tableofcontents
\endgroup

%==================================
% Introduction
\section{Introduction}\label{sec:intro}
%==================================

At the end of a quasi-de Sitter phase of inflation~\cite{Guth:1980zm,Linde:1981mu,Albrecht:1982wi,Starobinsky:1980te} there is a transition period, often referred to as reheating (see e.g., Refs.~\cite{Kolb:1990vq,Allahverdi:2010xz,Amin:2014eta,Baumann:2014nda,Sato:2015dga} for a review), that eventually leads to a universe dominated by relativistic particle degrees of freedom and the commencement of the radiation-dominated phase. In the context of slow-roll inflation this reheating occurs during coherent oscillations of a scalar field.  In the simplest models of single-field, slow-roll inflation the oscillating scalar field is the inflaton field itself. If the inflationary potential contains an inflection point, as in hilltop inflationary models (see e.g.~\cite{Boubekeur:2005zm}) and certain hybrid inflationary models (see e.g., \cite{Schmitz:2018nhb,Pallis:2013dxa,Bose:2013kya,Armillis:2012bs,Lin:2008ys,Kohri:2007gq}), then the oscillation frequency of the scalar field during coherent oscillations can typically be much larger than the Hubble expansion rate during inflation.  In such situations, gravitational particle production can occur without an exponential suppression even when the mass $m_\chi$ of the particles being gravitationally produced is much larger than the Hubble expansion rate $\Hinf$ during inflation \cite{Ema:2018ucl,Ema:2015dka,Ema:2016hlw}.  (A different strategy for super-Hubble mass production is presented in \rref{Hashiba:2018iff}.)

Unlike the case of parametric resonance~\cite{Dolgov:1989us,Traschen:1990sw,Shtanov:1994ce,Kofman:1997yn}, this particle production can be qualitatively interpreted perturbatively as a $2\rightarrow2$ scattering process \cite{Ema:2018ucl} where two cold inflaton $\phi$ particles produce two final-state $\chi$ particles through gravitational interactions. These particles in principle can serve as dark-matter particles if they are stable or sufficiently long lived. They can also, in principle, cause unwanted late-time decay problems. In \rref{Ema:2018ucl} the gravitational particle production was analytically estimated in the perturbative scattering picture, and numerically computed using the Bogoliubov formalism (i.e., computing the time evolution of the $\mathrm{SU}(1,1)$ rotation matrix). In the numerical calculation there are two classes of rates:  the fast frequencies of the oscillation frequency $\omega_*$ of the coherently oscillating field and that of the spectator field $\chi$ (typically $m_\chi$) and the slow frequency of the expansion rate of the universe, $H$.  Because of the disparate timescales of the fast and slow frequencies the numerical system is stiff, and the oscillatory nature of the system makes the brute-force  numerical integration difficult.  Hence, an analytic formalism is desirable.

\begin{figure}[t]
\begin{center}
\includegraphics[width=0.95\textwidth]{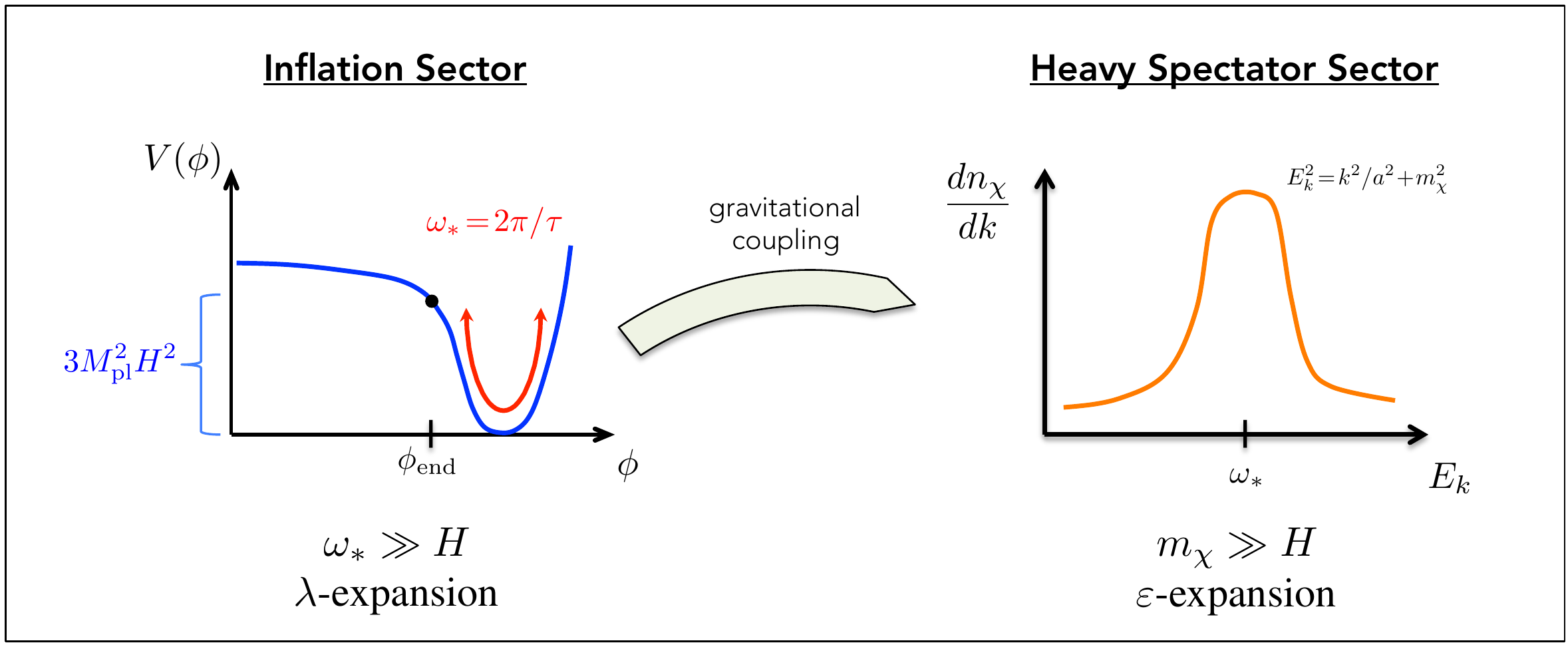} 
\caption{\label{fig:cartoon}
Rapid oscillations of the inflaton field at the end of inflation lead to the production of heavy spectator particles via the coupling of each field to gravity.  In the regime where the inflaton field oscillates quickly, $\omega_\ast \gg H$, and the spectator particles are heavy, $m_\chi \gg H$, we show that an the particle spectrum can be calculated analytically using the Bogoliubov formalism.  }
\end{center}
\end{figure}

In this paper, we provide such an analytic formalism to compute the spectrum of particles that are gravitationally produced in the regime where the background geometry has a rapidly-oscillating component, $\omega_\ast \gg H$, and where the spectator field is heavy, $m_\chi \gg H$.  An analytical calculation is made possible by the double expansion in the small ratios $\lambda \sim H / \omega_\ast \ll 1$ and $\varepsilon \sim H / m_\chi \ll 1$.  We provide a graphical representation of the essential ingredients of our calculation in \fref{fig:cartoon}.  

The formalism applies for a canonically normalized single real scalar field $\phi$ with a potential $V(\phi)$ dominating the coherent oscillations period. We do not deal with the inhomogeneities of the inflaton condensate in the inflated Hubble patch and focus on the homogeneous condensate effects. Any inhomogeneous effects are expected to be complementary to the subject of this paper as long as parametric resonance effects are unimportant. 

As an application of this formalism we consider the model of gravitational particle production presented in \rref{Ema:2018ucl}.  That reference evaluated the number density of $\chi$ particles, which are produced in the first few Hubble times after time $t_{\rm{end}}$ marking the end of inflation, with a qualitative perturbative calculation, finding 
\begin{align}
	n_\chi(t) & = \Cc H_\mathrm{inf}^3 \frac{m_\chi^4}{m_\phi^4} 
\ \Theta(m_\phi - m_\chi)\left(\frac{a(t_{\rm end})}{a(t)} \right)^3 \qquad & \text{(conformally-coupled scalar)} \\ 
	n_\chi(t) & = \Cm H_\mathrm{inf}^3 \ \Theta(m_\phi - m_\chi) \left(\frac{a(t_{\rm end})}{a(t)} \right)^3\qquad & \text{(minimally-coupled scalar)} 
\end{align}
where the coefficients were estimated using numerical techniques.  Here we show that $\Cc = (3/32)(1/16\pi) \simeq 0.002$ and $\Cm = (3/8)(1/16\pi) \simeq 0.007$ and compute  mass dependent factors.  We also provide a more general analytic formula applicable to the case of arbitrary gravitational coupling $\xi$ and larger range of $m_\chi$ and $m_\phi$ masses.

The order of presentation is as follows. In Sec.~\ref{sec:GPP} we briefly review the formalism of particle production in an expanding universe.  In Sec.~\ref{sec:exponential} we discuss the origin of the usual exponential suppression for $m_\chi> H$. In Sec.~\ref{sec:oscillation} we develop a general, model-independent formalism for evaluating particle production by expressing the Bogoliubov coefficient $\beta_k$ in terms of the Fourier transforms of the Hubble parameter and the Ricci scalar.  In Sec.~\ref{sec:slow_fast_breakdown} we present details of the decomposition of the energy-momentum density and the evolution of the background geometry into fast and slow components. Sec.~\ref{sec:General-formalism} summarizes the simple computational procedure of the formalism explicitly. In Sec.~\ref{sec:Application}, we apply the procedure of Sec.~\ref{sec:General-formalism} to a simple phenomenologically relevant quadratic coherent oscillation potential. Sec.~\ref{sec:Conclusions} concludes the paper.

%==================================
% Particle production in an expanding universe
\section{\label{sec:GPP}Particle production in an expanding universe}
%==================================

We begin by briefly reviewing the machinery behind gravitational particle production.  We work in the Friedmann-Lema\^itre-Roberston-Walker (FLRW) spacetime; its metric can be written as 
\begin{align}
	\ud s^2 = g_{\mu\nu} \ud x^\mu \ud x^\nu = \ud t^2 - a^2(t) \, \abs{\ud \xvec}^2 = a^2(\eta)\left( \ud \eta^2 - \abs{\ud \xvec}^2 \right)
\end{align}
where the scale factor $a$ is a function of a temporal coordinate: either coordinate time $t$ or conformal time $\eta$.  The Hubble parameter $H$ and Ricci scalar $R$ are given by 
\begin{subequations}\label{eq:H_and_R}
\begin{align}
	H & = \frac{\dot{a}}{a} = \frac{a^\prime}{a^2} \\ 
	R & = - 6 \, \frac{\ddot{a}}{a} - 6 \, \frac{\dot{a}^2}{a^2} = - 6 \, \frac{a^{\prime\prime}}{a^3}
\end{align}
\end{subequations}
where $\dot{a} = \partial_t a$ and $a^\prime = \partial_\eta a$.  The geometry responds to the energy-momentum tensor $T^{\mu\nu}$, which is assumed to arise primarily from a nearly homogeneous scalar field $\phi(t)$ with canonically-normalized kinetic term.  In the $(t,\vec{x})$ coordinates that we defined above, the energy momentum tensor has the usual energy density and pressure components  $T^{\mu}_{\phantom{\mu}\nu} = (\rho+P) \, \delta^{\mu}_{\phantom{\mu}0} \delta^{0}_{\phantom{\mu}\nu}-  P \delta^\mu_{\phantom{\mu}\nu} $ with 
\begin{align}\label{eq:rho_and_P}
	\rho = \frac{1}{2} \dot{\phi}^2 + V(\phi)
	\qquad \text{and} \qquad 
	P = \frac{1}{2} \dot{\phi}^2 - V(\phi)
	\per
\end{align}
Einstein's equations give 
\begin{align}\label{eq:Einstein_eqn}
	3 \Mpl^2 H^2 = \rho 
	\qquad \text{and} \qquad 
	- \Mpl^2 R = T^{\mu\nu} g_{\mu\nu}  =T = \rho - 3 P
\end{align}
where $\Mpl = (8 \pi G_N)^{-1/2} \simeq 2.43 \times 10^{18} \GeV$ is the reduced Planck mass.

Consider a real scalar field $\chi(x)$ with mass $m_\chi$ that is non-minimally coupled to gravity with coupling parameter $\xi$.  The Fourier modes of this field satisfy a mode equation, $\chi_k^{\prime\prime} + \omega_k^2 \chi_k = 0$, where the dispersion relation is given by 
\begin{align}\label{eq:omega_k}
	\omega_k = \sqrt{k^2 + a^2 m_\chi^2 + \Cxi a^2 R} 
	\com
\end{align}
and $k = \abs{\kvec}$ is the magnitude of the comoving wavevector.  The Fourier modes can be broken up into Bogoliubov coefficients, $\alpha_k$ and $\beta_k$, and the mode equation becomes 
\begin{subequations}
\begin{align}
	\alpha_k^\prime & = \frac{1}{2} \, N_k \, \beta_k \, e^{i \Phi_k} \\ 
	\beta_k^\prime  & = \frac{1}{2} \, N_k \, \alpha_k \, e^{-i \Phi_k} 
\end{align}
\end{subequations}
where the time-dependent phase is defined by
\begin{align}
	\Phi_k 
	\equiv 2 \int_{\eta_1}^{\eta} \! \ud \eta^\prime \, \omega_k(\eta^\prime) 
	= 2 \int_{t_1}^{t} \! \ud t^\prime \, \frac{\omega_k(t^\prime)}{a(t^\prime)}
	\com
\end{align}
where $\eta_1$ and $t_1$ set the phase $\Phi_k(\eta_1)=\Phi_k(t_1)=0$, and the non-adiabaticity parameter $N_k$ is defined by $N_k \equiv \omega^\prime_k / \omega_k$ or equivalently $N_k / a = \dot{\omega}_k / \omega_k$.  
If gravitational particle production is a small perturbation on the field, then $\alpha_k \approx 1$ and 
\begin{align}
	\beta_k(\eta) = \beta_k(\eta_P) + \frac{1}{2} \int_{\eta_P}^\eta \! \ud \eta^\prime \, N_k(\eta^\prime) \, e^{-i \Phi_k(\eta^\prime)} 
	\per
\end{align}
Note that $t_1 \geq t_P$ fixes the phase convention, and $t_P$ partially controls the adiabaticity with which one-particle states are defined.  
If the field $\chi$ is initially in its vacuum state then $|\beta_k(\eta_P)| \ll |\beta_k(\eta)|$, and we have 
\begin{align}
\label{eq:mainbogo-1}
	\beta_k = \int_{\eta_P}^\eta \! \ud \eta^\prime \ \frac{\omega_k^\prime}{2\omega_k} \ \mathrm{exp}\left[ -2i \int_{\eta_1}^{\eta^\prime} \! \ud \eta^{\prime\prime} \, \omega_k \right] 
	\per
\end{align}
Using the dispersion relation \pref{eq:omega_k} and moving to coordinate time gives the master equation 
\begin{align}\label{eq:beta_k_1}
	\beta_k
	= \frac{1}{2} \int_{t_P}^t \! \ud t^\prime \, \left\{\frac{H(t^\prime) m_\chi^2 + \Cxi \left[ \frac{1}{2} \dot{R}(t^\prime) + H(t^\prime) R(t^\prime) \right]}{\left[ \omega_k(t^\prime)/a(t^\prime) \right]^{2} }\right\} \, \mathrm{exp}\left[ -2i \int_{t_1}^{t^\prime} \! \ud t^{\prime\prime} \, \frac{\omega_k(t^{\prime\prime})}{a(t^{\prime\prime})} \right] \com
\end{align}
where the term in curly brackets is simply $\dot\omega_k/\omega_k$.  Assuming that the non-gravitational interactions of $\chi$ can be neglected, in terms of $\beta_k$ the spectrum of gravitationally-produced $\chi$-particles is given by 
\begin{align}\label{eq:dn_chi}
	\ud n_\chi 
	= \frac{\ud^3 \kvec}{(2\pi)^3} \, \lim_{t\to \infty} \frac{1}{a^3} \, \abs{\beta_k}^2 
	= \frac{k^3 \, \ud \! \ln k}{2\pi^2} \, \lim_{t\to \infty} \frac{1}{a^3} \, \abs{\beta_k}^2 
	\per
\end{align}
Integrating over the spectrum gives the total $\chi$-particle density, and the relic abundance, $\Omega_\chi = m_\chi n_\chi / 3 \Mpl^2 H_0^2$, given by 
\begin{align}\label{eq:Omega_chi_h2}
	\Omega_\chi h^2 & \simeq \left( 0.0271 \right) \left( \frac{m_\chi}{10^{10} \GeV} \right) \left( \frac{n_\chi / s_0}{10^{-20}} \right)
\end{align}
where $H_0 = 100 h \km / \mathrm{sec} / \mathrm{Mpc}$, $s_0 = (2\pi^2/45) g_{\ast S,0} T_0^3$, $g_{\ast S,0} \simeq 3.91$, and $T_0 \simeq 0.234 \meV$.  
Thus, the typical procedure is to solve for $H(t)$, $R(t)$, and $a(t)$ through Einstein's equations and $\phi(t)$ equations of motion.  Afterwards, one uses Eqs.~(\ref{eq:beta_k_1}) and (\ref{eq:dn_chi}) to compute the spectator $\chi$ particles that are produced, where the neglect of $\chi$ particle production back reactions are often well justified away from the parametric resonance regime.

With this information in hand, the integral in \eref{eq:beta_k_1} can be performed to evaluate $\beta_k(t)$ and the spectrum is given by \eref{eq:dn_chi}.

%==================================
% Heavy particles and exponentially suppression
\section{\label{sec:exponential}Heavy particles and exponential suppression}
%==================================

In many models of inflation and reheating, gravitational production of heavy particles leads to an exponentially-suppressed abundance (see for example \cite{Chung:1998bt,Birrell:1982ix}).  In this context, ``heavy'' particle species have masses that are larger than the Hubble scale at the time of particle production.  For such models we can write 
\begin{equation}\label{eq:generic_expectation}
	n_{\chi} \propto \mathrm{exp}[ -c_\chi m_\chi / H ]
\end{equation}
where $c_\chi$ is a model-dependent coefficient. Evidently gravitational particle production is negligible for $m_\chi \gg H$.  However, the generic expectation in \eref {eq:generic_expectation} is not universally applicable, and in the remainder of this section we discuss a class of models in which the exponential suppression is evaded.  

Let us first understand how the exponential suppression in \eref {eq:generic_expectation} arises generically.  From \eref{eq:dn_chi} we recall that $n_\chi \sim \abs{\beta_k}^2$, and from \eref{eq:beta_k_1} we recall that $\beta_k \sim \int \! \ud t^\prime \, (\dot{\omega}_k/\omega_k) \times \mathrm{[phase\ factor]}$.  In the regime of interest for heavy particles, $m_\chi \gg H$, the phase factor is simply $\mathrm{exp}[ - 2 i \omega_k t^\prime]$.  If $\dot{\omega}_k / \omega_k$ is slowly varying then the time integral over a rapidly-oscillating phase leads to a small $\abs{\beta_k}$.  To put it another way, the integral over $t^\prime$ is a Fourier transform that selects out oscillations on the time scale $\tau \sim (2 \omega_k)^{-1}$, but if $\dot{\omega}_k / \omega_k$ only varies on a much longer time scale, $\tau \sim H^{-1}$, then the Fourier transform gives $\beta_k \sim \mathrm{exp}[-b_\chi \omega_k /(a H)]$ with $b_\chi = O(1)$.  Such models lead to the exponential suppression observed in \eref{eq:generic_expectation}.  

However, the would-be exponential suppression can be avoided in models of inflation and reheating for which $\omega_k$ also contains a rapidly-oscillating component.  Suppose that $\omega_k$ oscillates with a period $\tau$ such that $m_\chi^{-1} \lesssim \tau/2\pi \ll H^{-1}$. Then oscillations in $\dot{\omega}_k / \omega_k$ are cancelled by the phase factor in \eref{eq:beta_k_1} when $2\pi/\tau \approx (k^2 / a^2 + m_\chi^2)^{1/2}$.  In other words, the Fourier transform picks out the rapidly-oscillating mode, which does not have an exponentially-suppressed amplitude.  For models in which the exponential suppression is avoided, one can also view gravitational particle production as a result of $2\rightarrow2$ scattering~\cite{Ema:2018ucl}.  

A rapidly-oscillating dispersion relation arises in many interesting models of inflation and reheating.  Coherent oscillations of the inflaton sector during reheating (this includes the waterfall field in the case of hybrid inflation) lead to an oscillating energy density, which induces oscillations in the scale factor $a(t)$ through Einstein's equations, and consequently the spectator field's dispersion relation, $\omega_k$, also acquires an oscillating component.  In the following sections we present a general formalism for studying gravitational particle production in such models.

%==================================
% Particle production from background oscillations
\section{\label{sec:oscillation}Particle production from background oscillations}
%==================================

We have argued in \sref{sec:exponential} that the relic abundance of gravitationally-produced particles may evade the generic exponential suppression for $m_\chi \gg H$ provided that the dispersion relation, $\omega_k$, has a rapidly-oscillating component.  For such models, it is generally difficult to evaluate the relic abundance with numerical techniques, since the integrand of \eref{eq:beta_k_1} is rapidly oscillating.  In this section we develop a more general, model-independent formalism for evaluating the relic abundance analytically.  By doing so we show that the Bogoliubov coefficient $\beta_k$ can be expressed simply in terms of the Fourier transform of the Hubble parameter and Ricci scalar.  

%---------------------------------------------------
% Assumption 1: \quad small-fast much larger than large-slow
\subsection{Assumption \#1: \ \ \ small-fast much larger than large-slow}
%---------------------------------------------------

We are interested in a system for which the energy-momentum tensor can be written as the sum of two terms:  the first term is large in magnitude and evolves slowly, whereas the second term is small in magnitude and evolves (oscillates) quickly.  This is the situation, for instance, in many models of reheating where the inflaton energy density is slowly redshifting while a sub-dominant component is also rapidly oscillating.  Thus, we decompose the energy-momentum tensor as 
\begin{align}\label{eq:T_mu_nu}
	T^{\mu\nu} & = T^{\mu\nu}_\slow + \lambda \, T^{\mu\nu}_\fast
\end{align}
where the dimensionless variable $\lambda$ will be used to perform a formal expansion in the ratio $\mathrm{[small~\&~fast]} / \mathrm{[large~\&~slow]}$, and we will eventually take $\lambda \to 1$ at the end of the calculation.  We provide additional details of the slow-fast decomposition in \sref{sec:slow_fast_breakdown}.  There, we will explain how this is an expansion in the asymptotic limit of $H/\omega_*$ which can be accomplished by taking $\Mpl\rightarrow \infty$ with the energy-momentum tensor fixed.

Einstein's equations \pref{eq:Einstein_eqn} describe the response of the spacetime metric to the energy-momentum tensor.  Thus we can also decompose the FLRW scale factor, the Hubble parameter, and the Ricci scalar into large-slow and small-fast components.  In summary we will write 
\begin{subequations}\label{eq:slow_fast}
\begin{align}
	a & = a_\slow + \lambda \, a_\fast + O(\lambda^2) \\ 
	H & = H_\slow + \lambda \, H_\fast + O(\lambda^2) \\ 
	R & = R_\slow + \lambda \, R_\fast 
	\per
\end{align}
\end{subequations}
The relation between $R$ and $T^{\mu\nu}$ is linear, but the relations with $a$ and $H$ are nonlinear.  

Using the slow/fast decomposition in \eref{eq:slow_fast}, we evaluate the Bogoliubov coefficient $\beta_k$ from \eref{eq:beta_k_1}.  We perform a formal expansion in powers of $\lambda$ to obtain 
\begin{align}\label{eq:beta_k_2}
	\beta_k
	= \frac{1}{2} \int_{t_P}^t \! \ud t^\prime \,  & \left\{
	\left[ \frac{H_\slow m_\chi^2 + \Cxi \left( \frac{1}{2} \dot{R}_\slow + H_\slow R_\slow \right)}{\left( k^2/a_\slow^2 + m_\chi^2 + \Cxi R_\slow \right)} \right] \right.
	\\ & 
	+ \left[ 
	\frac{k^2}{a_\slow^2} \left( \Cxi \dot{R}_\slow + 2 H_\slow m_\chi^2 + \frac{1}{3} (1-6\xi) H_\slow R_\slow \right) \frac{a_\fast}{a_\slow} \right.
\nn & \qquad 
	+ \left( m_\chi^2 + \Cxi R_\slow \right) \left( \frac{k^2}{a_\slow^2} + m_\chi^2 + \Cxi R_\slow \right) H_\fast
	\nn & \qquad 
	+ \Cxi \left( H_\slow \frac{k^2}{a_\slow^2} - \frac{1}{2} \Cxi \dot{R}_\slow \right) R_\fast
	\nn & \qquad 
	\left. + \frac{1}{12} (1-6\xi) \left( \frac{k^2}{a_\slow^2} + m_\chi^2 + \Cxi R_\slow \right) \dot{R}_\fast \right]
	\nn & \quad 
	\left. \times \frac{\lambda}{\left( k^2/a_\slow^2 + m_\chi^2 + \Cxi R_\slow \right)^2} + O(\lambda^2) \right\} \, \mathrm{exp}\left[ -2i \int_{t_1}^{t^\prime} \! \ud t^{\prime\prime} \, \frac{\omega_k(t^{\prime\prime})}{a(t^{\prime\prime})} \right] \nonumber
	\per
\end{align}
The terms of $O(\lambda^2)$ are negligible as long as the fast terms are small in magnitude as compared to the slow terms; this assumption is justified and clarified in \sref{sec:slow_fast_breakdown}.  

%---------------------------------------------------
% Assumption \#2: \ \ \ \ \ $H$ much less than $m_\chi$
\subsection{Assumption \#2: \ \ \  Expansion rate much less than mass of spectator field}
%---------------------------------------------------

We are interested in systems for which the Hubble expansion rate is small compared to the mass of the spectator scalar field, $H \ll m_\chi$, and the spectator particles are heavy.  If we employ the slow-fast decomposition from \eref{eq:slow_fast}, we can write $H_\fast \ll H_\slow \approx H \ll m_\chi$, and therefore the small expansion parameter is $H_\slow / m_\chi$.  

How do $R_\slow$ and $\dot{R}_\slow$ compare with $m_\chi$?  We assume that the Hubble expansion rate controls the time rate of change of the slow components.  In other words, $\dot{a}_\slow \sim H_\slow a_\slow$, $\ddot{a}_\slow \sim H_\slow^2 a_\slow$, $R_\slow \sim H_\slow^2$, and $\dot{R}_\slow \sim H_\slow R_\slow$.  Therefore, in order to keep track of the expansion in $H_\slow / m_\chi \ll 1$ we replace
\begin{align}\label{eq:epsilon_expansion}
	m_\chi \to \varepsilon^0 \, m_\chi
	\, , \quad 
	H_\slow \to \varepsilon^1 \, H_\slow
	\, , \quad 
	R_\slow \to \varepsilon^2 \, R_\slow
	\, , \quad 
	\dot{R}_\slow \to \varepsilon^3 \, \dot{R}_\slow 
	\per
\end{align}
At the end of the calculation we will take $\varepsilon \to 1$ again.  

Using the replacements in \eref{eq:epsilon_expansion} we evaluate the Bogoliubov coefficient from \eref{eq:beta_k_2}.  We perform an expansion in small $\varepsilon$ to find 
\begin{align}\label{eq:beta_k_3}
	\beta_k
	= \frac{1}{2} \int_{t_P}^t \! \ud t^\prime \, & \left\{ 
	\left[ H_\slow \frac{m_\chi^2}{E_k^2} \, \varepsilon + {\cal O} (\varepsilon^2) \right] 
	\ + \ \left[ 
	\frac{k^2}{a_\slow^2} (2H_\slow m_\chi^2) \frac{a_\fast}{a_\slow} \ \varepsilon 
	+ m_\chi^2 E_k^2 H_\fast \right. \right.
	\nn & \ \ \
	+ \left. \Cxi H_\slow \frac{k^2}{a_\slow^2} R_\fast \ \varepsilon 
	+ \frac{1}{12} (1-6\xi) E_k^2 \dot{R}_\fast
	+ O(\varepsilon^2) \right] \frac{\lambda}{E_k^4}
	\nn & \ \ \ 
	\left. + \, O(\lambda^2) \phantom{\frac{a^2}{m^2}} \hspace*{-16pt}\right\}  \, \exp\left[ -2i \int_{t_1}^{t^\prime} \! \ud t^{\prime\prime} \, E_k(t^{\prime\prime}) + O(\varepsilon^2) \right] 
\com
\end{align}
where we have also defined $E_k(t) \equiv \sqrt{k^2 / a^2_\slow(t) + m_\chi^2}$.  

%---------------------------------------------------
% Further simplifications
\subsection{Further simplifications}
%---------------------------------------------------

Subject to the assumptions of the previous subsections, the Bogoliubov coefficient $\beta_k$ is given by \eref{eq:beta_k_3}.  We now retain only the leading-order terms to obtain 
\begin{align}\label{eq:beta_k_4}
	\beta_k
	= \int_{t_P}^t \! \ud t^\prime \, \left\{ 
	\frac{H_\slow}{2}
	+ \frac{H_\fast}{2}
	+ \frac{1}{24} (1-6\xi) \frac{\dot{R}_\fast}{m_\chi^2}
	\right\} \, \frac{m_\chi^2}{E_k^2} \ \mathrm{exp}\left[ -2i \int_{t_1}^{t^\prime} \! \ud t^{\prime\prime} \, E_k(t^{\prime\prime}) \right] 
\end{align}
where we have also set $\lambda = \varepsilon = 1$. This expression can be simplified further.  

%==========
\paragraph*{Term with $H_\slow$.  }
We now argue that the first term in $\beta_k$ can be neglected.  Recall that $H_\slow$ only varies on a time scale $\Delta t \sim O(H_\slow^{-1})$.  However, on this time scale, the phase factor is rapidly oscillating, because $\Delta t \sim E_k^{-1} < m_\chi^{-1} \ll H_\slow^{-1}$.  Therefore the first term in $\beta_k$ integrates down to zero when we send $t \to \infty$, and it can be neglected.  

%==========
\paragraph*{Term with $\dot{R}_\fast$.  }
The third term involves a time derivative, $\dot{R}_\fast$.  We can evaluate the integral by parts.  The boundary terms vanish, and the term that goes as $\dot{E}_k \sim H_\slow E_k$ can be neglected.  Thus we are left with 
\begin{align}\label{eq:beta_k_5}
	\beta_k(t)
	= \int_{t_P}^t \! \ud t^\prime \, \left\{ 
	\frac{H_\fast(t^\prime)}{2}
	+ \frac{i}{12} (1-6\xi) \frac{R_\fast(t^\prime) \, E_k(t^\prime)}{m_\chi^2}
	\right\} \, \frac{m_\chi^2}{E_k(t^\prime)^2} \, \mathrm{exp}\left[ -2i \int_{t_1}^{t^\prime} \! \ud t^{\prime\prime} \, E_k(t^{\prime\prime}) \right] 
	\per
\end{align}

%==========
\paragraph*{Time scales.  }
In practice the rapidly-oscillating contributions to the Hubble parameter and Ricci scalar are not present for all times, but they only appear after inflation has ended and the inflaton sector experiences coherent oscillations about the minimum of its potential.  Thus we expect most of the gravitational particle production to occur just after the end of the quasi-dS phase of inflation at time $\tend$, and it is convenient to take $t_1 = \tend$.  If the fast components oscillate with a period $\tau$ then the Fourier transform will pick out modes with $E_k \sim 2\pi/\tau \equiv \omega_\ast$, and since we are interested in fast oscillations $\omega_\ast  \gg H$ it follows that the integral will be dominated by $t^\prime \sim \tend + O(E_k^{-1})$, since the phase factor is rapidly oscillating on larger time scales.  Since $E_k(t)$ only changes on the much longer time scale, $\Delta t \sim O(H^{-1})$, we can approximate $E_k(t) \approx E_k(\tend)$ and evaluate the integral in the exponent.  Since we know that the $t^\prime$ integral is dominated by times $(t^\prime - \tend) \ll H^{-1}$, we can extend the limits of integration to infinity.  After these manipulations we have 
\begin{align}\label{eq:beta_k_6}
	\beta_k
	= \int_{-\infty}^\infty \! \ud t^\prime \, \left\{ 
	\frac{H_\fast(t^\prime)}{2}
	+ \frac{i}{12} (1-6\xi) \frac{R_\fast(t^\prime) \, E_k(\tend)}{m_\chi^2}
	\right\} \, \frac{m_\chi^2}{E_k^2(\tend)} \, \mathrm{exp}\left[ -2i E_k(\tend) \, (t^\prime - \tend) \right] 
	\per
\end{align}

%==========
\paragraph*{Identify Fourier transform.  }
The remaining time integral now takes the form of a Fourier transform.  We define the Fourier transform by 
\begin{align}
	H_\mathrm{fast}(t) & = \int_{-\infty}^{\infty} \! \frac{\ud \omega}{2\pi} \, \tilde{H}_\mathrm{fast}(\omega) \, e^{-i \omega t} 
	\qquad \text{and} \qquad 
	\tilde{H}_\mathrm{fast}(\omega) = \int_{-\infty}^{\infty} \! \ud t^\prime \, H_\mathrm{fast}(t^\prime) \, e^{i \omega t^\prime} \\
	R_\mathrm{fast}(t) & = \int_{-\infty}^{\infty} \! \frac{\ud \omega}{2\pi} \, \tilde{R}_\mathrm{fast}(\omega) \, e^{-i \omega t} 
	\qquad \text{and} \qquad 
	\tilde{R}_\mathrm{fast}(\omega) = \int_{-\infty}^{\infty} \! \ud t^\prime \, R_\mathrm{fast}(t^\prime) \, e^{i \omega t^\prime} 
	\com
\end{align}
and the expression for $\beta_k$ in \eref{eq:beta_k_6} becomes simply 
\begin{align}\label{eq:beta_k_7}
	\beta_k
	= \frac{1}{2} \tilde{H}_\fast \frac{m_\chi^2}{E_k^2} 
	+ \frac{i}{12} (1-6\xi) \, \tilde{R}_\fast \frac{1}{E_k} 
\end{align}
where $\tilde{H}_\fast(\omega)$ and $\tilde{R}_\fast(\omega)$ are evaluated at $\omega = - 2 E_k$, and $E_k(t)$ is evaluated at $t = \tend$.  
This formula is a main result of our paper; it illustrates how fast oscillations in a sub-leading contribution to the Hubble parameter and Ricci scalar leads to gravitational particle production.  

%==========
\paragraph*{Apply Einstein's equations.  }
The Hubble parameter and Ricci scalar acquire components which are rapidly oscillating components because they respond to a rapidly-oscillating energy-momentum tensor through Einstein's equations \pref{eq:Einstein_eqn}.  We will make this mapping explicit in \sref{sub:slow_fast_geometry}, but let us now anticipate that result here by writing
\begin{align}\label{eq:beta_k_8}
	\beta_k
	= \frac{1}{4 \sqrt{3} \Mpl} \frac{\tilde{\rho}_\fast}{\sqrt{\rho_\slow}} \frac{m_\chi^2}{E_k^2} 
	- i \, \frac{1-6\xi}{12\Mpl^2} \, \left[ \tilde{\rho}_\fast - 3 \tilde{P}_\fast \right] \frac{1}{E_k} 
\end{align}
where $\tilde{\rho}_\fast(\omega)$ and $\tilde{P}_\fast(\omega)$ are evaluated at $\omega = - 2 E_k$, and $E_k(t)$ and $\rho_\slow(t)$ are evaluated at $t = \tend$.  It is remarkable that the evaluation of $\beta_k$ reduces to simply calculating the Fourier transforms of the energy density and pressure.

%==================================
% Details of the slow-fast decomposition
\section{\label{sec:slow_fast_breakdown}Details of the slow-fast decomposition}
%==================================

In the previous section we derived \eref{eq:beta_k_7}, which expresses the Bogoliubov coefficient $\beta_k$ in terms of $\tilde{H}_\fast$ and $\tilde{R}_\fast$, which are the Fourier transforms of the rapidly-oscillating parts of the Hubble parameter and the Ricci scalar.  We begin this section by defining the slow-fast decomposition more precisely.  We also derive expressions for $\tilde{H}_\fast$ and $\tilde{R}_\fast$ that relates them to the energy density and pressure, $\tilde{\rho}_\fast$ and $\tilde{P}_\fast$, which justifies the formula for $\beta_k$ in \eref{eq:beta_k_8}.  Next we impose energy-momentum conservation to solve for the time-dependence of $\rho_\fast(t)$ and $P_\fast(t)$, and we evaluate their Fourier transforms.  Finally we assume a Gaussian spectral model to evaluate the Fourier coefficients that appear in \eref{eq:beta_k_8}.  

%---------------------------------------------------
% Slow-fast decomposition of the energy-momentum tensor
\subsection{\label{sub:slow_fast_stress_energy}Slow-fast decomposition of the energy-momentum tensor}
%---------------------------------------------------

Previously in \eref{eq:T_mu_nu} we wrote $T^{\mu\nu} = T_\slow^{\mu\nu} + \lambda \, T_\fast^{\mu\nu}$, which assumes that the energy-momentum tensor can be decomposed into a slowly-evolving term and a rapidly-oscillating term.  Let us now provide more precise definitions of $T_\slow^{\mu\nu}$ and $T_\fast^{\mu\nu}$. This is done in the following way. The energy-momentum tensor admits a Fourier decomposition, 
\begin{align}
	T^{\mu\nu}(t) = \int_{-\infty}^{\infty} \frac{\ud \omega}{2\pi} \, \tilde{T}^{\mu\nu}(\omega) \, e^{- i \omega t} 
	\qquad \text{where} \qquad 
	\tilde{T}^{\mu\nu}(\omega) = \left[ \tilde{T}^{\mu\nu}(-\omega) \right].
\end{align}
To quantitatively separate the ``fast modes'' which have large $\omega$ and the ``slow modes'' which have small $\omega$, we introduce a fiducial separation scale, $\Lambda_\IR$, which satisfies $\Lambda_\IR \gg H$, and define 
\begin{align}\label{eq:T_slow_fast}
	T^{\mu\nu} = T_\slow^{\mu\nu} + \lambda \, T_\fast^{\mu\nu} 
	\qquad \text{with} \qquad 
	\begin{cases}
	T_\slow^{\mu\nu}(t) = \mathlarger{\int}_{-\Lambda_\IR}^{\Lambda_\IR} \dfrac{\ud \omega}{2\pi} \, \tilde{T}^{\mu\nu}(\omega) \, e^{- i \omega t} \\[6pt] 
	T_\fast^{\mu\nu}(t) = \left( \mathlarger{\int}_{-\infty}^{-\Lambda_\IR} + \mathlarger{\int}_{\Lambda_\IR}^{\infty} \right) \dfrac{\ud \omega}{2\pi} \, \tilde{T}^{\mu\nu}(\omega) \, e^{- i \omega t} 
	\end{cases}
	\per
\end{align}
Now, $T_\slow^{\mu\nu}$ encodes the time dependence of the slow modes and $T_\fast^{\mu\nu}$ represents the fast modes.  Ideally, $\tilde{T}^{\mu\nu}(\omega)$ is bimodal having a peak at $|\omega| \ll \Lambda_\IR$ corresponding to the slow modes, and a second peak at $|\omega| \gg \Lambda_\IR$ corresponding to the fast modes, and in this case the distinction between slow and fast is insensitive to the arbitrary choice of $\Lambda_\IR$.  

With the definition of \eref{eq:rho_and_P}, the energy density and pressure inherit the slow-fast decomposition of \eref{eq:T_slow_fast}:  
\begin{subequations}\label{eq:rho_P_slow_fast}
\begin{align}
	\rho & = \rho_\slow + \lambda \, \rho_\fast \\ 
	P & = P_\slow + \lambda \, P_\fast 
	\per
\end{align}
\end{subequations}

Previously in \sref{sec:oscillation} we have assumed that $\abs{T_\fast^{\mu\nu}} \ll \abs{T_\slow^{\mu\nu}}$.   As we will discuss further in \sref{sub:expansion_param}, the origin of this hierarchy is the smallness of the rate at which energy conservation is violated due to the expansion of the universe compared to the quasi-periodic oscillation time scale associated with the field potential.  In the limit of exact energy conservation (no expansion), the fast component is negligible since the energy density is all in the slowest mode, i.e., constant energy density.  After we make the formal expansion in $\lambda$, we will check the self-consistency of the assumption in \eref{eq:lamerror}. 

%---------------------------------------------------
% Slow-fast decomposition of the background geometry
\subsection{\label{sub:slow_fast_geometry}Slow-fast decomposition of the background geometry}
%---------------------------------------------------

Once a slow-fast decomposition is performed for the energy-momentum tensor $T^{\mu\nu}$, then slow-fast decompositions for the Ricci scalar $R$, Hubble parameter $H$, and scale factor $a$ are derived.  
Einstein's equation \pref{eq:Einstein_eqn} provides a linear relation between $R$ and $T^{\mu\nu}$, and we find 
\begin{align}\label{eq:R_slow_fast}
	R = R_\slow + \lambda \, R_\fast 
	\qquad \text{with} \qquad 
	\begin{cases}
	R_\slow = - \dfrac{\rho_\slow - 3 P_\slow}{\Mpl^2} \\[6pt] 
	R_\fast = - \dfrac{\rho_\fast - 3 P_\fast}{\Mpl^2} 
	\end{cases}
\end{align}
where $\rho$ and $P$ admit slow-fast decompositions via \eref{eq:T_slow_fast}.  
The relation between $H$ and $T^{\mu\nu}$ is nonlinear \pref {eq:Einstein_eqn}, but nevertheless $H$ admits a slow-fast decomposition as a series expansion in powers of $\lambda$, which gives 
\begin{align}\label{eq:H_slow_fast}
	H = H_\slow + \lambda \, H_\fast + O(\lambda^2)
	\qquad \text{with} \qquad 
	\begin{cases}
	H_\slow = \dfrac{\sqrt{\rho_\slow}}{\sqrt{3} \Mpl} \\[6pt]
	H_\fast = \dfrac{1}{2} \dfrac{1}{\sqrt{3} \Mpl} \dfrac{\rho_\fast}{\sqrt{\rho_\slow}} 
	\end{cases} 
	\per
\end{align}
Finally, the relation between $a$ and $H$ is also nonlinear \pref{eq:H_and_R}, 
\begin{align}
	\log \left[ \frac{a(t)}{a(\tend)} \right] 
	& = \int_{\tend}^t \! \ud t^\prime \, \left[ H_\slow(t^\prime) + \lambda \, H_\fast(t^\prime) \right] \com
\end{align}
but if we focus on short time scales, $(t-\tend) \ll H_\slow^{-1}$, then we can write
\begin{align}
	\log \left[ 1 + \left( \frac{a_\slow(t)}{a(\tend)} - 1 \right) + \lambda \, \frac{a_\fast(t)}{a(\tend)} \right] 
	& = H_\slow(\tend) \left( t - \tend \right) + \lambda \int_{\tend}^t \! \ud t^\prime \, H_\fast(t^\prime) 
	\per
\end{align}
We expand in powers of $\lambda$ and match terms to find 
\begin{align}\label{eq:a_slow_fast}
	a = a_\slow + \lambda \, a_\fast + O(\lambda^2)
	\qquad \text{with} \qquad 
	\begin{cases}
	\dfrac{a_\slow(t)}{a(\tend)} = 1 + \dfrac{\sqrt{\rho_\slow(\tend)}}{\sqrt{3} \Mpl} \, \left( t-\tend \right) \\[6pt] 
	\dfrac{a_\fast(t)}{a(\tend)} = \dfrac{1}{2} \dfrac{1}{\sqrt{3} \Mpl}  \dfrac{\mathlarger{\int}_{\tend}^t \! \ud t^\prime \,\rho_\fast(t^\prime)}{\sqrt{\rho_\slow(\tend)}} 
	\end{cases}
\end{align}
where we have also used the expressions for $H_\slow$ and $H_\fast$ from \eref{eq:H_slow_fast}.  Since $\rho_\slow$ appears in the denominator of the expression for $a_\fast$, the power counting in $\Mpl$ depends on whether
we keep $H_\slow$ or $\rho_\slow$ fixed as we take $\Mpl \to \infty$.

%---------------------------------------------------
% Evolution from energy-momentum conservation
\subsection{\label{sub:energy_momentum}Evolution from energy-momentum conservation}
%---------------------------------------------------

In this section we will use energy-momentum conservation to derive expressions for $\rho_\fast(t)$ and $P_\fast(t)$, which will allow us to evaluate their Fourier transforms in the next section.  

The energy-momentum conservation condition, $\nabla_{\mu} T^{\mu\nu} = 0$, gives 
\begin{align}\label{eq:energy_conservation}
	\dot{\rho} + 3H \left( \rho+P \right) = 0 
	\per
\end{align}
We can evaluate this formula using the slow-fast decompositions from Secs.~\ref{sub:slow_fast_stress_energy}~and~\ref{sub:slow_fast_geometry}.  
At $O(\lambda^0)$ we obtain 
\begin{align}\label{eq:d_rho_slow}
	\dot{\rho}_\slow + 3 H_\slow \left( \rho_\slow + P_\slow \right) = 0 
	\com
\end{align}
which governs the evolution of $\rho_\slow$. At $O(\lambda^1)$ we obtain 
\begin{align}\label{eq:d_rho_fast}
	\dot{\rho}_\fast + 3 H_\slow \left( \rho_\fast + P_\fast \right) + \frac{3}{2} H_\slow \rho_\fast \left( 1 + \frac{P_\slow}{\rho_\slow} \right) = 0 
\end{align}
where we have also used \eref{eq:H_slow_fast} to write $H_\fast = H_\slow \rho_\fast / 2 \rho_\slow$.  

Next we seek to solve \eref{eq:d_rho_fast} for $\rho_\fast(t)$.  To evaluate $\rho_\fast + P_\fast$, we recall that the energy-momentum tensor is dominantly controlled by the evolution of a scalar field \pref{eq:rho_and_P}; this gives $\rho + P = \dot{\phi}^2$. As long as the frequency scale $\omega_*$ for the oscillation of $\phi$ is much larger than $H$, we can neglect the Hubble expansion rate in describing $\dot{\phi}^2$ time evolution on a time scale smaller than $H^{-1}$.  This means if we denote $\phiC$ as the solution to the equation of motion that neglects $H$ (i.e.~an energy conserving solution to $\ddot{\phi} + 3 H \dot{\phi} + V^{\prime}(\phi) = 0$ with $H=0$), then $\dot{\phi}^2$ will be dominated by the $\omega_*$ scale oscillation behavior without any slow time evolution coming from $H$.  This means we can approximate $\dot{\phi}^2 \approx ( \dot{\phi}^2 )_\fast$ on a time scale smaller than $H^{-1}$, allowing us to conclude
\begin{align}\label{eq:rho_plus_P}
	\rho_\fast + P_\fast = \bigl( \dot{\phi}^2 \bigr)_\fast \approx \dphiC^2 \left(1 + O(\lambda) \right)
\end{align}
where the identification of neglecting $H$ with the $\lambda$ expansion that appears on the right hand side will be justified later by imposing that result obtained from $( \dot{\phi}^2 )_\fast \approx \dphiC^2$ approximation be self-consistent with the $\lambda$ expansion that we have been using (see \eref{eq:limits}).  Although one can in principle solve for $\phiC(t)$ implicitly in terms of integrals, it is not very useful for computing the desired Fourier integral that are needed to evaluate $\beta_k$.  

Now \eref{eq:d_rho_fast} is solved for $\rho_\fast(t)$ by direct integration.  If we focus on short time scales, $(t-\tend) \ll H^{-1}$, then the slow parameters can be treated as constants and we find  
\begin{align}\label{eq:rho_fast}
	\rho_\fast(t) & =  \rho_\fast(\tend)\ \exp\left[ -\frac{3}{2} H_\slow (1+P_\slow / \rho_\slow) (t-\tend) \right] - 3 H_\slow \int_{\tend}^t \! \ud t^\prime \, \dphiC^2(t^\prime) \nonumber \\
& \approx \rho_\fast(\tend) - 3 H_\slow \int_{\tend}^t \! \ud t^\prime \, \dphiC^2(t^\prime) \com
\end{align}
where $H_\slow$, $\rho_\slow$, and $P_\slow$ are evaluated at $t = \tend$.  
The first term is approximately constant, which does not belong in $\rho_\fast$, but it is cancelled by the second term with a judicious choice of $\tend$. 
Using \eref{eq:rho_plus_P} we find 
\begin{align}\label{eq:P_fast}
	P_\fast(t) = \dphiC^2(t) - 3 H_\slow \int_{\tend}^t \! \ud t^\prime \, \dphiC^2(t^\prime)
\end{align}
where $H_\slow = H_\slow(\tend)$.  Hence, we see that the oscillatory time scale of $\rho_\fast(t)$ and $P_\fast(t)$ is set by the period $\tau$ of $\phiC(t)$. (We will give an explicit expression for $\tau$ in terms of the potential in \eref{eq:omega_ast}.)

%---------------------------------------------------
% Understanding the expansion parameters
\subsection{\label{sub:expansion_param}Understanding the expansion parameters}
%---------------------------------------------------

Let us now pause to discuss briefly the nature of the book-keeping variable $\lambda$ that we have been using to keep track of the $\mathrm{[small~\&~fast]} / \mathrm{[large~\&~slow]}$ expansion.  Using the expression for $\rho_\fast$ from \eref{eq:rho_fast} we can check the self-consistency of the $\lambda$ expansion.  Consider the ratio 
\begin{align}\label{eq:lamerror}
	\frac{|\rho_\fast(t)|}{\rho_\slow(t)}  
	\ = \  3H_\slow \left| \int_{\tend}^{t} \ud t^\prime \,\frac{\frac{1}{2} \dphiC^{2}(t^\prime)}{\rho_\slow(t)} \right|
	\ \lesssim \  3 H_\slow \tau \com
\end{align}
where $H_\slow = H_\slow(\tend)$.  
The integrand is no larger than one, since $\rho$ contains $\dot{\phi}^2/2$.  
Moreover, the integral does not grow without bound like $\sim (t-\tend)$, because the integrand also oscillates, and then the integral is largest when it equals $\tau$, the period of $\phiC$. Thus we understand that powers of $\lambda$ are keeping track of powers of the small quantity, $H_\slow \tau$:  
\begin{equation}\label{eq:self-consistent-interpret} 
	O([H_\slow \tau]^{n}) = \text{kept track of by $\lambda^n$ terms}
	\per
\end{equation}
The formal expansion in powers of $\lambda$ is a reliable approximation as long as $\phiC$ oscillates on a time scale that is short compared to the Hubble time.  

We can further develop our understanding of the $\lambda$ expansion by inspecting \eref{eq:d_rho_slow} and \eref{eq:d_rho_fast}.  Suppose that we send $H_\slow \to 0$.  Since $H_\slow$ and $\rho_\slow$ are related by the Friedmann equation, which is $3 \Mpl^2 H_\slow^2 = \rho_\slow$ at $O(\lambda^0)$, the limit $H_\slow \to 0$ also corresponds to $\Mpl \to \infty$ with $\rho_\slow$ held fixed.  Taking $H_\slow \to 0$ enforces $\dot{\rho}_\slow \to 0$, which is realized when $\rho_\slow$ is approximately constant and energy is conserved, but it also enforces $\dot{\rho}_\fast \to 0$, which implies a small amplitude for $\rho_\fast$, since the rapidly-oscillating term $\rho_\fast$ cannot be constant by definition.  In other words, the expansion in $\mathrm{[small~\&~fast]} / \mathrm{[large~\&~slow]}$ that we have been parametrizing with the variable $\lambda$ can also be viewed as the limit in which energy is conserved:  
\begin{align}\label{eq:limits}
	\left[ \lambda \to 0 \right] 
	\quad \Longleftrightarrow \quad 
	\left[ H_\slow \to 0 \right] 
	\quad \Longleftrightarrow \quad 
	\left[ \begin{array}{c}\Mpl \to \infty \\ \rho_\slow = \const \end{array} \right] 
	\quad \Longleftrightarrow \quad 
	\left[ \rho_\fast \to 0  \right] 
	\per
\end{align}
Illuminating these connections is a main result of this study.  

%---------------------------------------------------
% Fourier transforms of energy density and pressure
\subsection{\label{sub:Fourier_trans}Fourier transforms of energy density and pressure}
%---------------------------------------------------

In order to evaluate $\beta_k$ using the expression in \eref{eq:beta_k_8} we must first evaluate $\tilde{\rho}_\fast(\omega)$ and $\tilde{P}_\fast(\omega)$, which are the Fourier transforms of $\rho_\fast(t)$ and $P_\fast(t)$.  
Using the expression for $\rho_\fast(t)$ from \eref{eq:rho_fast}, it is straightforward to evaluate the Fourier transform, which gives
\begin{align}\label{eq:tilde_rho_fast}
	\tilde{\rho}_\fast(\omega) \approx -3i H_\slow(\tend) \int_{-\infty}^{\infty} \! \frac{\ud \omega^\prime}{2\pi} \frac{\omega^\prime \, (\omega-\omega^\prime)}{\omega} \, \tphiC(\omega^\prime) \, \tphiC(\omega-\omega^\prime)
\per
\end{align}
Similarly we could evaluate $\tilde{P}_\fast(\omega)$ by taking the Fourier transform of $P_\fast(t)$ from \eref{eq:P_fast}, but a more useful expression is obtained by taking the transform of \eref{eq:d_rho_fast} instead to find 
\begin{align}
	-i \omega \tilde{\rho}_\fast + 3 H_\slow \left( \tilde{\rho}_\fast + \tilde{P}_\fast \right) + \frac{3}{2} H_\slow \tilde{\rho}_\fast \left( 1 + w \right) = 0 
	\com
\end{align}
and solving for $\tilde{P}_\fast$ gives 
\begin{align}\label{eq:rho_minus_3P}
	\tilde{\rho}_\fast - 3 \tilde{P}_\fast = \left( - \frac{i\omega}{H_\slow} + \frac{11}{2} + \frac{3}{2} \frac{P_\slow}{\rho_\slow} \right)\tilde{\rho}_\fast	
\end{align}
where $H_\slow$, $\rho_\slow$, and $P_\slow$ are evaluated at $t = \tend$.  
Note that for $\omega \gg H_\slow$ we can neglect the term containing $w = P_\slow / \rho_\slow$, since $P_\slow$ is always smaller than $\rho_\slow$.  
This says that the particle production is insensitive to the equation of state of the slowly evolving part of the background spacetime to leading order in $\lambda$.

%---------------------------------------------------
% Gaussian spectral model
\subsection{\label{sub:Gaussian}Gaussian spectral model}
%---------------------------------------------------

Further evaluation of $\tilde{\rho}_\fast(\omega)$ from \eref{eq:tilde_rho_fast} requires an expression for $\tphiC(\omega)$.  By construction $\phiC(t)$ is periodic and smooth, since it corresponds to the energy-conserving solution.  Therefore we now adopt the Gaussian spectral model, 
\begin{align}\label{eq:tilde_phi_C}
	\tphiC(\omega) = \frac{A}{\sqrt{2\pi\sigma^{2}}} \left[ e^{-(\omega-\omega_\ast)^{2}/(2\sigma^{2})} + e^{-(\omega+\omega_\ast)^{2}/(2\sigma^{2})}\right] e^{-i\omega \tend} 
	\per
\end{align}
Note that $\tphiC(\omega) = \tphiC^\ast(-\omega)$ such that $\phiC(t) = \phiC^\ast(t)$.  

The three parameters $\omega_\ast$, $A$, and $\sigma$ can be extracted from the scalar potential, $V(\phi)$.  The dominant mode has a frequency $\omega_\ast = 2 \pi / \tau$, where $\tau$ is the period of oscillation, $\phiC(\tend) = \phiC(\tend + \tau)$.  The energy-conservation constraint implies 
\begin{align}\label{eq:omega_ast}
	\omega_\ast \equiv \pi \left(\int_{\phiC(\tend)}^{\phiC(\tend+\frac{\tau}{2})}\frac{\ud\phi}{\sqrt{2V(\phiC(\tend))-2V(\phi)}}\right)^{-1}
\end{align}
where we have assumed $V(\phiC(\tend+\tau/2))=V(\phiC(\tend))$ and $\phi(\tend+\tau/2)>\phi(\tend)$. The amplitude $A$ is calculated as 
\begin{align}
	\phiC(\tend) 
	= \frac{A}{\sqrt{2\pi\sigma^{2}}} \int \! \frac{\ud\omega}{2\pi}\left[e^{-(\omega-\omega_\ast)^{2}/(2\sigma^{2})}+e^{-(\omega+\omega_\ast)^{2}/(2\sigma^{2})}\right]
	= \frac{A}{\pi} 
	\per
\end{align}
The variance is calculated as 
\begin{align}\label{eq:sigma_sq}
	\sigma^{2}
	& \approx - \omega_\ast^2 + \frac{\int \! \dfrac{d\omega}{2\pi} \, \tphiC(\omega) \, \omega^2 \, e^{i\omega \tend} }{\phiC(\tend)}
	= - \omega_\ast^2 + \frac{V'(\phiC(\tend))}{\phiC(\tend)} 
	\per
\end{align}

We are now equipped to evaluate $\tilde{\rho}_\fast(\omega)$ from \eref{eq:tilde_rho_fast}.  Employing the Gaussian spectral model gives 
\begin{align}\label{eq:mainresult} 
	\tilde{\rho}_\fast(\omega) 
	& \approx \frac{-3i \sqrt{\pi}}{16} \frac{H_\slow(\tend) \phiC^{2}(\tend)}{\sqrt{\sigma^{2}}} \frac{\left(\omega^{2}-2\sigma^{2}\right)}{\omega}
	\left( e^{-(\omega + 2 \omega_\ast)^2/4\sigma^2} + e^{-(\omega - 2 \omega_\ast)^2/4\sigma^2} \right) \, e^{-i\omega \tend}
 \per
\end{align}
Note that we have dropped the ``cross term,'' which contains a factor of $e^{-\omega^2 / 4 \sigma^2}$, because it does not contribute to fast modes.  

It is illuminating to investigate this expression in the limit $\sigma \to 0$.  We find 
\begin{align}
	\tilde{\rho}_\fast(\omega)
	\approx -\frac{3i \pi}{2^{7/2}} \, H_\slow(\tend) \, \phiC^{2}(\tend) \, \omega \, \left(\delta\left[\frac{\omega+2\omega_\ast}{\sqrt{2}}\right]+\delta\left[\frac{\omega-2\omega_\ast}{\sqrt{2}}\right]\right) \, e^{-i\omega \tend} 
	\com
\end{align}
and the inverse Fourier transform is 
\begin{align}\label{eq:sigma_to_zero}
	\rho_\fast(t) = \frac{3}{4} \, \omega_\ast \, H_\slow(\tend) \, \phiC^{2}(\tend) \, \sin[2\omega_\ast(t-\tend)]
\per
\end{align}
For comparison consider the model of a real scalar field with a harmonic potential, $V(\phi) = m_\phi^2 \phi^2 / 2$.  The equation of motion is solved by 
\begin{align}
	\phi(t) \approx \frac{\phiC(\tend)}{t/\tend} \, \cos[m_\phi(t-\tend)] \, \left\{ 1+O\left(H-\frac{2}{3t}\right)\right\} 
	\com
\end{align}
and the corresponding energy density is given by \eref{eq:sigma_to_zero} with $\omega_\ast = m_\phi$.  This gives us a nontrivial check of the Gaussian formalism for the case of the quadratic potential.

%==================================
% General formalism
\section{\label{sec:General-formalism}General formalism}
%==================================

Here we use the results of the previous sections to state a simple algorithm for computing the spectrum of heavy spectator particles that arise from gravitational particle production for a given inflaton potential $V(\phi)$ and in the limit $m_\chi/H(\tend)\gg1$ where $\tend$ is the time at the end of the quasi-dS era.  We assume that the energy-momentum tensor is dominated by a single real scalar field $\phi$ with a canonically normalized kinetic term.  

\paragraph*{1.}
Compute the peak frequency $\omega_\ast$.  
For a given potential $V(\phi)$ one can solve
\begin{equation}
	\frac{\Mpl^{2}}{2}\left(\frac{V'(\phi_\mathrm{end})}{V(\phi_\mathrm{end})}\right)^{2}=1
\end{equation}
for $\phi_\mathrm{end}$ to find the field value at the time period just after the end of the quasi-dS phase when $\epsilon_{V}(\tend)=1$.  
This provides an initial condition for the energy-conserving solution $\phiC(\tend) = \phi_\mathrm{end}$.  
Then one can solve 
\begin{equation}
	V\left( \phiC(\tend) \right)=V\left( \phiC(\tend+\tau/2) \right)
\end{equation}
for $\phiC(\tend + \tau/2)$ to find the turning points of the potential.  
Note that this definition of the turning point ignores the dissipation by Hubble friction as when constructing an adiabatic invariant, which is reasonable for $\omega_\ast \gg H(\tend)$.  Finally one can evaluate \eref{eq:omega_ast}: 
\begin{equation}
\label{eq:peakfreq}
	\omega_\ast \equiv \pi\left[\int_{\phiC(\tend)}^{\phiC(\tend+\frac{\tau}{2})}\frac{\ud\phi}{\sqrt{2V(\phiC(\tend))-2V(\phi)}}\right]^{-1} 
	\com
\end{equation}
which assumes $\phiC(\tend) < \phiC(\tend + \tau/2)$.  
If the integral can be done analytically, one does not have to explicitly compute $\phi_\mathrm{end}$.  If the integral cannot be computed analytically due to the complexities of the potential, one can fix $\phiC(\tend)$ and compute $\omega_\ast$.  However, it is important to recognize that it is far easier to numerically evaluate this integral (and work through the following procedure) than it is to perform the numerical integration in \eref{eq:beta_k_1} that's required to evaluate $\beta_k$ by ``brute force'' methods, because $H \ll m_\chi$.  

\paragraph*{2.}
Compute the Fourier transforms of the energy density and pressure, $\tilde{\rho}_\fast(\omega)$ and $\tilde{P}_\fast(\omega)$.  Assuming that the oscillations of $\phiC(t)$ can be described by the Gaussian spectral model, the energy density is given by \eref{eq:mainresult}:  
\begin{align}
	\tilde{\rho}_\fast(\omega) 
	& \approx \frac{3i \sqrt{\pi}}{16} \frac{H_\slow \phiC^2}{\sqrt{\sigma^{2}}} \frac{\omega^2 - 2\sigma^2}{\omega} \left( e^{-(\omega + 2 \omega_\ast)^2/4\sigma^2} + e^{-(\omega - 2 \omega_\ast)^2/4\sigma^2} \right) \, e^{i\omega \tend} 
	\com
\end{align}
and Friedmann's equation relates 
\begin{equation}
	\rho_\slow = 3 \Mpl^2 H_\slow^2 = V(\phiC) 
\end{equation}
where $\rho_\slow$, $H_\slow$, and $\phiC$ are evaluated at $\tend$ both here and below.  
The spectral variance is given by \eref{eq:sigma_sq}: 
\begin{equation}
	\sigma^{2}\equiv\frac{V'(\phiC)}{\phiC}-\omega_\ast^{2}
	\per
\end{equation}
The pressure is given by \eref{eq:rho_minus_3P}:
\begin{align}
	\tilde{\rho}_\fast - 3 \tilde{P}_\fast = \left( - \frac{i\omega}{H_\slow} + \frac{11}{2} \right)\tilde{\rho}_\fast 
	\com
\end{align}
where a term of order $ P_\slow / \rho_\slow < 11/2$ has been dropped inside the parantheses.  

\paragraph*{3.}
Compute the Bogoliubov coefficient $\beta_k$.  One can evaluate \eref{eq:beta_k_8}: 
\begin{align}
	\beta_k
	= \frac{1}{4 \sqrt{3} \Mpl} \frac{\tilde{\rho}_\fast}{\sqrt{\rho_\slow}} \frac{m_\chi^2}{E_k^2} 
	- i \, \frac{1-6\xi}{12\Mpl^2} \, \left[ \tilde{\rho}_\fast - 3 \tilde{P}_\fast \right] \frac{1}{E_k} 
\end{align}
where $\tilde{\rho}_\fast$ and $\tilde{P}_\fast$ are evaluated at $\omega = - 2E_k$, and $E_k(t)$ is evaluated at $\tend$: 
\begin{equation}
	E_k = \sqrt{\frac{k^2}{a_\slow^2(\tend)} + m_\chi^2} 
	\per
\end{equation}
One can interpret $k/a_\slow(\tend)$ as the free-particle momentum of the $\chi$ particle at time $\tend$.  If the field $\chi$ is conformally coupled to gravity ($\xi = 1/6$) then the second term in $\beta_k$ vanishes.  
Otherwise we have 
\begin{align}\label{eq:beta_k_ratio}
	\frac{\beta_k \ \text{[second term]}}{\beta_k \ \text{[first term]}} 
	\approx (1-6\xi) \left( \frac{2 E_k^2}{m_\chi^2} \right) 
\end{align}
where we have dropped a sub-leading term that is $O(H_\slow / E_k) < O(H_\slow / m_\chi) \ll 1$ in the regime of interest.  Particle production is most efficient for modes with $E_k \approx \omega_\ast$, and therefore modes with $E_k \gg m_\chi$ will dominate the phase space as long as $\omega_\ast \gg m_\chi$.  (This is equivalent to there being a hierarchy in the book-keeping expansion parameters, $\varepsilon \gg \lambda$). Consequently the second term in $\beta_k$ is generally larger than the first one, except when $\xi \approx 1/6$.  

%==========
\paragraph*{4.}
Compute the spectrum of  $\chi$-particle production rate.  
Most of the particle production occurs within the first few Hubble times after the oscillations begin, i.e times $t$ such that $O(\omega_\ast^{-1}) \ll t - \tend < O(H_\slow^{-1})$.  The rapid $\phiC$ oscillations that are driving particle production have their amplitude damped away on a time scale $H_\slow^{-1}$, and therefore particle production shuts off and the comoving spectrum is conserved at later times, i.e. $a^3(t) \, dn_\chi/dk |_t \approx \const$ for $t - \tend > O(H_\slow^{-1})$.  To determine the late time relic abundance of $\chi$ particles we can calculate the spectrum of particle production {\em rate} close to $t_{\rm end}$.  The assumptions that underlie our analytic result restrict its regime of validity to times $t \sim t_2$ such that $O(E_k^{-1}) \ll t_2 - \tend \ll O(H_\slow^{-1})$; the lower limit follows from validity of the Fourier transform in \eref{eq:beta_k_6}, and the upper limit follows from neglecting the damping of $\phiC$ in \eref{eq:rho_plus_P}.  We calculate the production rate spectrum during this time using \eref{eq:dn_chi}, which gives 
\begin{align}\label{eq:dndk_formalism}
	\frac{d \gamma(t_{\rm{end}})}{dk} \equiv \frac{1}{t_2-t_{\rm{end}}}\frac{d n_\chi}{dk} \Bigr|_{t_2}
	\approx  \frac{\sigma}{\sqrt{2\pi}}\frac{k^2}{2\pi^2} \, \frac{1}{a_\slow^3(t_{\rm end})} \, \abs{\beta_k(t_2)}^2 
	\qquad \text{for} \qquad 
	E_k^{-1} \ll t_2 - \tend \ll H_\slow^{-1}
.
\end{align}
If $E_{k}\gg|\sigma|$
(e.g.~close to quadratic potential), the expression simplifies as
\begin{equation}\label{eq:smallwidth}
  \frac{d\gamma(t_{\rm{end}})}{dE_{k}}  \approx \frac{(E_{k}^{2}-m_{\chi}^{2})^{1/2}E_{k}}{2048\pi} \left( \frac{m_{\chi}^{2}}{E_{k}^{2}}+2(1-6\xi)\right)^{\! 2} \left( \frac{\phiC}{\Mpl} \right)^{\! 4}\frac{E_{k}^{2}}{\sqrt{2\pi}\sigma} \, \exp\left[-\frac{2(E_{k}-\omega_{*})^{2}}{\sigma^{2}}\right] 
	\per
\end{equation}
This gives the production rate as a function of time $t_{\rm end}$, and with the replacement of the time parameter $t_{\rm end}\rightarrow t$, this  can be used in Boltzmann equations as a function of time $t$ to compute the particle production in a standard manner.\footnote{Such a Boltzmann equation is defined to be coarse grained on a time scale of $t_2-t_{\rm end}$.} 
This expression for the spectrum is an approximation in the limit that $E_{k}/H\gg1$ with an error budget of $\max[H/\omega_\ast,H/E_{k}]$.  
Another source of error in this computation is the Gaussian modeling of the frequency distribution.  Such approximations are invalid if the coherent oscillations contain many strong frequencies as might be the case for multi-field coherent oscillations with many oscillation scales.  Furthermore, since the natural correspondence is $\sigma^{-1}=(t_2-t_{\rm end})/ \sqrt{2\pi}$, one cannot rigorously apply the Gaussian ansatz for $\sigma^{-1} \gg  (t_2-t_{\rm end})/ \sqrt{2\pi}$, but as long as Fourier transform is a good approximation (which effectively treats $t_2-t_{\rm end}$ as an infinite time period), one can allow for even smaller widths.

%==================================
% Application to a quadratic potential
\section{\label{sec:Application}Application to a quadratic potential}
%==================================

To illustrate a simple and phenomenologically useful application of this formalism, we now consider a class of models that can be described by a quadratic inflaton  potential, 
\begin{equation}
	V(\phi) = \frac{1}{2} m_\phi^{2} \phi^{2} 
	\com
\end{equation}
near the local minimum at $\phi = 0$ where reheating occurs.  Away from $\phi = 0$ the quadratic potential may be extended (as in the hilltop inflation model of \rref{Ema:2018ucl}).  We use this potential to compute the number of $\chi$ particles that are gravitationally produced for $m_\chi\gg H$ by following the steps laid out in \sref{sec:General-formalism}.  

\paragraph*{1.}
To compute the peak frequency (\eref{eq:peakfreq}), we evaluate
\begin{align}
	\int_{\phi(\tend)}^{\phi(\tend+\frac{\tau}{2})} \frac{\ud\phi}{\sqrt{2V(\phiC(\tend))-2V(\phi)}}= \frac{\pi}{m_\phi} 
	\com
\end{align}
and the peak frequency is found to be
\begin{align}
	\omega_\ast = m_\phi 
	\per
\end{align}

\paragraph*{2.}
We compute the frequency dispersion, finding 
\begin{equation}
	\sigma^2 = 0 
	\per
\end{equation}
Consequently the Fourier transforms of the energy density and pressure are evaluated in the $\sigma^2 \to 0$ limit, which gives 
\begin{align}
	\tilde{\rho}_\fast
	& \approx \frac{3i \pi}{2^{7/2}} \, H_\slow \phiC^2 \omega \, \left(\delta\left[\frac{\omega+2m_\phi}{\sqrt{2}}\right]+\delta\left[\frac{\omega-2m_\phi}{\sqrt{2}}\right]\right) \, e^{i \omega \tend} \\ 
	\tilde{\rho}_\fast - 3 \tilde{P}_\fast 
	& \approx \left( - \frac{i\omega}{H_\slow} + \frac{11}{2} \right) \tilde{\rho}_\fast 
\per
\end{align}
Here and throughout the rest of this section, $H_\slow$, $\phiC$, $\tilde{\rho}_\fast$, and $\tilde{P}_\fast$ are evaluated at $\tend$.   Friedmann's equation gives 
\begin{equation}\label{eq:Hslow_to_phiC}
	\rho_\slow = 3 \Mpl^2 H_\slow^2 = \frac{1}{2} m_\phi^2 \phiC^2
	\per
\end{equation}

\paragraph*{3.}
The Bogoliubov coefficient is then given by 
\begin{align}\label{eq:beta_k_application}
	\beta_k
	& = \left[ -i \frac{\pi H_\slow \phiC}{16 m_\phi \Mpl} \, \sqrt{3} E_k \, \delta(-\sqrt{2}E_k+\sqrt{2}m_\phi) \, e^{-2i E_k \tend} \right] \, \frac{m_\chi^2}{E_k^2} 
	\nn & \quad 
	+ (1-6\xi) \left[ -i \frac{\pi}{8\sqrt{2}} \, \frac{E_k^3 \, \phiC^2}{m_\chi^2 \Mpl^2} \, \delta(-\sqrt{2}E_k+\sqrt{2}m_\phi) \, e^{-2i E_k \tend} \right] \, \frac{m_\chi^2}{E_k^2} 
\com 
\end{align}
where $E_k$ is evaluated at $\tend$.  In the second term we have assumed $H_\slow / E_k \ll 1$.  

%==========
\paragraph*{4. - Conformal Coupling:}
If the spectator is conformally coupled to gravity ($\xi = 1/6$) then 
\begin{align}\label{eq:betafirststep1} 
	\abs{ \beta_{k}(t_2) }^{2} 
	& \approx  \lim_{\Delta t \rightarrow \infty} \Delta t \frac{3\pi}{1024} \, \frac{H_\slow^2 \phiC^2}{\Mpl^2} \, \frac{m_\chi^4}{m_\phi^2 E_k^2} \, \delta(E_k - m_\phi) \com
\end{align}
where we have used $\delta(\sqrt{2} x)^2 = \lim_{y \to \infty} (y/4\pi) \, \delta(x)$.  The spectrum's rate of change is given by 
\begin{align}\label{eq:spectrum-1}
	\frac{d\gamma_{\xi=1/6}(t_{\rm end})}{dk} 
	= \frac{3}{2048 \pi} \, \frac{k^2}{a_\slow^3} \, \frac{H_\slow^2 \phiC^2}{\Mpl^2} \, \frac{m_\chi^4}{m_\phi^2 E_k^2} \, \delta(E_k - m_\phi) 
\com
\end{align}
where $a_\slow$ is evaluated at $\tend$.  Integrating the spectrum gives the density production rate
\begin{align}\label{eq:gamma_conformal}
	\gamma_{\xi=1/6}(t_{\rm end}) 
	= \frac{9}{1024\pi} \, H_\slow^4 \, \frac{m_\chi^4}{m_\phi^4} \, \sqrt{1-\frac{m_\chi^2}{m_\phi^2}} \ \Theta(m_\phi - m_\chi) 
	\per
\end{align}
We can integrate the Boltzmann equation using \eref{eq:gamma_conformal} with $\tend$ promoted to $t$ and $a_\slow(t) \propto t^{2/3}$.  
This gives at $t \gg t_{\rm{end}}+1/H$
\begin{align}
	n_{\chi}(t) 
	& = \Cc \, H_\slow^{3} \, \frac{m_\chi^4}{m_\phi^4} \, \sqrt{1-\frac{m_\chi^2}{m_\phi^2}} \ \Theta(m_\phi - m_\chi)\left(\frac{a(t_{\rm end})}{a(t)} \right)^3 \label{eq:n_chi_conformal}
	\\ 
	\Cc 
	& \equiv \left( \frac{3}{32} \right) \left( \frac{1}{16\pi} \right) 
	\simeq 0.0019 \label{eq:C_conformal} 
	\per
\end{align}
The main result in \eref{eq:n_chi_conformal} gives the cosmological density of $\chi$ particles that arise from gravitational particle production due to the rapid oscillations of the inflaton field at the end of inflation; the formula assumes $H_\slow \ll m_\chi $ and that $\chi$ is conformally coupled to gravity ($\xi = 1/6$).  The volume dilution scaling assumes that the $\chi$ particles are free from particle-number-changing interactions at later times.  Note that this analytical calculation of $\Cc$ is consistent with the numerically-estimated coefficient of Eq.~(34) in \rref{Ema:2018ucl}.  Due to our approximations, the error on \eref{eq:n_chi_conformal} is expected to be of order $H_\slow(\tend)/m_\chi$.

\paragraph*{4. - Minimal Coupling:}
If the spectator is minimally coupled to gravity ($\xi = 0$), then both terms in \eref{eq:beta_k_application} contribute to $\beta_k$, but the second is larger by a factor of $2 E_k^2 / m_\chi^2 = 2 m_\phi^2 / m_\chi^2$.  Keeping only the second term, we follow the same calculations as before to now find 
\begin{align}
	 \frac{d\gamma_{\xi=0}(t_{\rm{end}})}{dk}  & = \frac{1}{1024\pi} \frac{k^2}{a_\slow^3} \frac{E_k^2 \, \phiC^4}{\Mpl^4} \, \delta(E_k - m_\phi) \\ 
	\gamma_{\xi=0}(t_{\rm{end}}) & = \frac{9}{256\pi} \, H_\slow^4 \, \sqrt{1 - \frac{m_\chi^2}{m_\phi^2}} \ \Theta(m_\phi - m_\chi) 
	\\
	n_{\chi}(t) 
	& = \Cm \, H_\slow^{3} \sqrt{1-\frac{m_\chi^{2}}{m_\phi^{2}}} \ \Theta(m_\phi - m_\chi)\left(\frac{a(t_{\rm end})}{a(t)} \right)^3 \label{eq:n_chi_minimal}
	\\ 
	\Cm 
	& = \left(\frac{3}{8}\right)\left(\frac{1}{16\pi}\right) 
	\simeq 0.0075 \label{eq:C_minimal} 
	\per
\end{align}
Equation~(\ref{eq:n_chi_minimal}) valid for $t\gg t_{\rm{end}}+1/H$ is another main result of this paper.  
The derivation of this result has assumed $H_\slow \ll m_\chi \ll m_\phi$ and that $\chi$ is minimally coupled to gravity ($\xi = 0$).  As $m_\chi$ approaches $m_\phi$, the two terms in the expression for $\beta_k$ from \eref{eq:beta_k_application} become comparable in magnitude, and this is why \eref{eq:n_chi_minimal} is only reliable for $m_\chi \ll m_\phi$.  Thus the threshold factor $(1 - m_\chi^2 / m_\phi^2)^{1/2} \approx 1$ must be close to one, but we have retained this factor anyway to illustrate that our formalism, i.e. the double expansion in $\lambda$ and $\varepsilon$, allows the computation of such terms.  It is remarkable that the analytical computation of $\Cm$ matches the numerical estimate in Eq.~(44) of \cite{Ema:2018ucl} very well.  The error in $\Cm$ is expected to be ${\cal O}(H_\slow / m_\chi)$.  

\paragraph*{4. - General Coupling and Masses:}  Keeping both terms in \eref{eq:beta_k_application} or using \eref{eq:smallwidth} we can also compute 
\begin{equation}
	n_{\chi}(t) =  \frac{3}{32} \, \frac{1}{16\pi} \, H_\slow^3(t_{\rm{end}})  \left( \frac{m_{\chi}^{2}}{m_{\phi}^{2}}+2(1-6\xi) \right)^{2} \sqrt{1-\frac{m_{\chi}^{2}}{m_{\phi}^{2}}} \ \Theta(m_{\phi}-m_{\chi})\left(\frac{a(t_{\rm end})}{a(t)} \right)^3
\label{eq:generalcoupling}
\end{equation}
for $t\gg t_{\rm{end}}+1/H$ where unlike in \eref{eq:n_chi_minimal} we have not made the $m_\chi \ll m_\phi$ approximation.

%==================================
% Conclusions
\section{\label{sec:Conclusions}Conclusions}
%==================================

In this work, we have presented a formalism to compute the scalar
particle $\chi$ spectrum produced by gravitational interactions of
the inflaton during the coherent oscillations in the kinematic regime
in which $m_\chi\gg H$. Unlike the parametric resonance case, this
is qualitatively the limit of $2\rightarrow2$ scattering through
gravitational interactions. 

The most general formula in terms of relevant Fourier components is given in \eref{eq:beta_k_8}. Next, we used the vanishing divergence of the energy-momentum tensor and a Gaussian approximation to the coherent oscillation frequency spectrum of a real canonically normalized scalar field to compute the relevant Fourier components. The result is in terms of the properties of the potential and the value of the inflaton at the end of inflation. The computational procedure is summarized in \sref{sec:General-formalism} and a  phenomenologically relevant quadratic coherent oscillation potential case is worked out in \sref{sec:Application}.  In the process, a compact small spectral width limit formula for the particle production rate spectrum is also presented in \eref{eq:smallwidth}.

One can apply the formalism in this paper to dark matter production in scenarios similar to \cite{Ema:2018ucl} as well as to moduli problem considerations in the spirit of \cite{Addazi:2017ulg,Evans:2013nka,Coughlan:1983ci}.  It would also be interesting to understand how different features in the coherent oscillation potential (such as cubic terms and radiative corrections) will change the gravitational production parametric dependence. The formulae could also be generalizable to higher-spin particle production. Finally, it would be interesting to systematically expand beyond a single Gaussian peak distribution model to more complicated multifield coherent oscillation frequency structure.

%----------------------------------------------------------------
% Acknowledgements
%----------------------------------------------------------------
\subsubsection*{Acknowledgements}
D.J.H.C.\ was supported in part by the DOE through grant DE-SC0017647. E.W.K.\ was supported in part by the DOE.  A.J.L.\ was supported at the University of Michigan by the US Department of Energy under grant DE-SC0007859. This work was performed in part at the Aspen Center for Physics, which is supported by National Science Foundation grant PHY-1066293. 

%----------------------------------------------------------------
% References
%----------------------------------------------------------------
\bibliographystyle{JHEP2}
\bibliography{super_Hub_mass}

\end{document}